\documentclass[12pt,a4paper,final]{article}
\usepackage[T1]{fontenc}
\usepackage[utf8]{inputenc}
\usepackage{authblk}
\usepackage{amsmath,amssymb,epsfig,cite,dsfont}
\usepackage{array,multirow}
\usepackage {color}
\usepackage[toc,page]{appendix}
\numberwithin{equation}{section}
\usepackage{afterpage}
\usepackage{cite}
\usepackage[notref,notcite]{showkeys}
\usepackage[numbers,sort&compress]{natbib}

\definecolor{verde}{cmyk}{.83,.21,1,.08}

\usepackage[colorlinks=true
,urlcolor=blue
,anchorcolor=blue
,citecolor=blue
,filecolor=blue
,linkcolor=blue
,menucolor=blue
,pagecolor=blue
,linktocpage=true
,pdfproducer=medialab
,pdfa=true
]{hyperref}

\advance\hoffset by -1.1truecm
\advance\voffset by -1.0truecm
\newcommand{\be}{\begin{equation}}
\newcommand{\ee}{\end{equation}}
\newcommand{\bea}{\begin{eqnarray}}
\newcommand{\eea}{\end{eqnarray}}

\newcommand{\del}{\partial}
\newcommand{\dd}{\mathrm d}
\newcommand{\ii}{\mathrm i}
\usepackage{mathrsfs}
\usepackage{authblk}

\numberwithin{equation}{section}
\usepackage{float}
\restylefloat{figure}
\newcounter{appendice}

\begin{document}

\title{ 
\begin{center} 
\bf{\Huge   Entangled Scent of a Charge
}\end{center}
}
\author[1]{M.  Asorey\thanks{asorey@unizar.es}}
\author[2]{A. P.~Balachandran\thanks{balachandran38@gmail.com}}
\author[3,4,5] {F. Lizzi\thanks{fedele.lizzi@na.infn.it}}
\author[3,4]{ G.~Marmo,\thanks{marmo@na.infn.it}}
\affil[1]{\small Departamento de F\'{\i}sica Te\'orica, 
Universidad de Zaragoza,  E-50009 Zaragoza, Spain.}
\affil[2]{\small Physics Department, Syracuse University, Syracuse, New York 13244-1130, U.S.A.}
\affil[3]{\small  Dipartimento di Fisica ``E. Pancini'' Universit\`a di
Napoli {\sl Federico II},  I-80125 Napoli, Italy. }
\affil[4]{\small  INFN Sezione di
Napoli,  I-80125 Napoli, Italy. }
\affil[5]{\small Departament de F\'{\i}sica Qu\`antica i Astrof\'{\i}sica and Institut de C\'{\i}encies del Cosmos (ICCUB),
Universitat de Barcelona. Barcelona, E-08007, Spain}

\renewcommand\Authands{ and }

\date{} 
\maketitle


\begin{abstract}
We argue that the ground state of a field theory, in the presence of charged particles, becomes an entangled state involving an infinity of soft photons. The quantum field {\sl vacuum} is altered by the  passage of a uniformly moving charge, leaving in its wake a different dressed ground state. In this sense a charged particle leaves its electromagnetic {\sl scent} even after passing by. Unlike in classical electrodynamics the effect of the charge remains even at infinite time.
The calculation is done in detail for the ground state of a spacetime wedge, although the results are more general. 
This agrees in spirit with recent results over the infrared aspects of field theory, although the technical details are different. These considerations open the possibility that the information carried by quantum fields, being nonlocal, does not disappear beyond the horizon of black holes.
\end{abstract}
\newpage

\section{Introduction}

Recently there is a growing perception of the importance of the infrared frontier in quantum field theory (for a texbook introduction see Weinberg or Nair~\cite{Weinbergvol2, Nair}), especially because of its deep connections with both gravity and information theory. For a recent review see for example~\cite{StromingerLectures} and its references. In particular, 
the cloud of infrared photon quantum states created from the vacuum by a uniformly moving charged particle belongs to a different superselection sector which is parametrized by  the linear momentum of the particle.  These different ground states can be distinguished by the action of a formally unitary operator related to coherent states. While the interest of original literature~\cite{Eriksson,Kibble1, Kibble2, Kibble3, Kibble4, FaddeevKulish,Zwanziger} was mostly in the S-matrix of quantum electrodynamics, recently the interest has shifted to the gravitational sector~\cite{Christodoulou, BieriGarfinkle}, and in particular to the issue of the infrared {gravitational}  memory and the information paradox~{\cite{GomezPanchenko,HawkingStromingerPerry, GomezLetschka, Carneyetal}}. In this paper we will consider  the gauge theory framework, emphasizing some quantum aspects which have several analogies with the gravitational results.

In a previous paper \cite{EOMASCON}, we related equations of motion and Gauss law constraints. 
In particular we analyzed  what happens in a limited region of spacetime, {e.g  a Rindler wedge, as discussed in Ref.~\cite{AsoBalMarQue}.}
In this paper we pursue this line of thought and argue that the superselected ground state structure depends, unlike  in classical electrodynamics, on the full trajectory of the particles, including also the portions of it which are outside (but still causally connected).  In this respect we will see that a particle that leaves a region  leaves its ``scent'' in its infrared structure.

\section{The ground states structure: infrared dressing}

 The main point of this section is to recall that the ground states of quantum electrodynamics have a very rich structure: Besides the Fock vacuum, there are an infinity of states with energies close to zero obtained by a limiting process of a growing number of infrared photons, with extremely low energies\footnote{{Although often in the literature these states are called vacua (without inverted commas), this is a misnomer. There is a unique quantum electrodynamics vacuum, the state annihilated by all annihilation operators. There can however be degenerate ground states in different superselection sectors. We are concerned with these states in this paper, and from now on we will refer to them as ground states.}}. The ground states which emerge are different, and a superselection rule separates them. This effect is of course not present in the quantum mechanics of a finite number of particles. This phenomenon has been known for a long time~\cite{Eriksson, Kibble1, Kibble2, Kibble3, Kibble4}, {for more recent discussion see also~\cite{Rasmussen}}.

Dirac \cite{Dirac} {(see also Roepstorff \cite{Roepstorff})} gave an elegant construction of the infrared dressed  charge. We will paraphrase the part of his work which concerns us as follows. Consider the smearing of the electromagnetic gauge field $A_\mu$ by \emph{test differential forms} $\varphi^\mu$ which we take to be smooth and vanishing at infinity. The gauge condition is imposed \emph{on test differential forms} by requiring that the forms $\varphi$ are co-closed: $d^*\varphi=\del_\mu \varphi^\mu=0$. We can then define 
\be
A(\varphi)=\int \dd^4x A_\mu(x) \varphi^\mu(x) \label{smearA},
\ee
which is a gauge invariant observable. This enables the construction of a Weyl system $\mathcal W$, which provides an exponentiated form of the commutation relations. Given two co-closed one-forms $\varphi_1$ and $\varphi_2$, we have
 \begin{equation}
\label{weyl}
{\mathcal W}(\varphi_1) {\mathcal W}(\varphi_2)= {\mathcal W}(\varphi_1+\varphi_2) e^{\ii \sigma(\varphi_1,\varphi_2) },
\end{equation}
where the {bilinear binary form} is given by
\be
\sigma(\varphi_1,\varphi_2)=\frac12\int \dd^4x\,\dd^4y  \varphi_1(x)^\mu D(x-y) \varphi_2(y)_\mu,
\ee
$D$ being the causal Pauli-Jordan {propagator}:
\begin{equation}
 D(x-y) = \int d\mu(\mathbf{k}) [e^{-\ii k\cdot(x-y)}-e^{\ii k\cdot(x-y)}]. \label{causalprog}
\end{equation}
The momentum space measure is as usual
\begin{equation}
 d\mu(\mathbf{k})= \frac{d^3 \mathbf{k}}{(2\pi)^3 2 k_0},  \label{measure}
\end{equation}
with $k_0=\sqrt{\mathbf{k}^2}=|\mathbf k|$. 
The causal  function $D$ satisfies the wave equation
\begin{equation}
\label{boxD}
 \square\, D(x)=0.
\end{equation}

The two-form $\sigma$ has a large kernel, it actually depends not on the one-forms $\varphi$'s, but on their ``curvature'',  i.e.\ only on the quantities 
\be
\varphi_{\mu\nu}(x)=\int \dd^4y D(x-y) (\del_\mu \varphi_\nu-\del_\nu \varphi_\nu)(y)
\ee
which are  solutions of the wave equation.

A representation of the Weyl system is obtained by the map
 \begin{equation}
\label{weyl2}
{\mathcal W}:\varphi\longrightarrow {\mathcal W}(\varphi)=e^{\ii A(\varphi)} .
\end{equation}
with $A_\mu$ acting on a Fock space.
But this is not the only representation.

Suppose that $\Phi$ is a real linear functional on the space of test forms:
 \begin{equation}
\label{line}
\Phi: \varphi\longrightarrow \Phi(\varphi) .
\end{equation}
Then the map 
 \begin{equation}
\label{weyl3}
{\mathcal W'}:\varphi\longrightarrow{\mathcal W}'(\varphi)= e^{\ii[A(\varphi)+\Phi(\varphi)\mathbb{I}]} 
\end{equation}
can equally well represent $\mathcal{W}$.

Given the mode expansion of $A_\mu$, 
\begin{equation}
\label{modes}
A_\mu(x)=\int d\mu(\mathbf{k})[\mathbf a_\mu (\mathbf{k}) e^{-i k\cdot x} + \mathbf a_\mu (\mathbf{k})^\dagger e^{\ii k\cdot x}],
\end{equation}
in terms of  the usual creation and annihilation operators $\mathbf a_\mu(\mathbf k)$ and $\mathbf a_\mu(\mathbf k)^\dagger$, 
and that of the test form $\varphi$,
\begin{equation}
\label{modes}
\tilde\varphi_\mu(\mathbf{k})= \int {d^4 {x}}\, \varphi_\mu({x}) e^{i  k\cdot   x},\quad k_0=|\mathbf{k}|,
\end{equation}
the gauge invariant observable (\ref{smearA}) is given by
\begin{equation}
\label{modes}
A(\varphi)=\int d\mu(\mathbf{k})\,[\tilde\varphi^\mu(\mathbf{k})^\ast \mathbf a_\mu  (\mathbf{k})
+ \tilde\varphi^\mu(\mathbf{k})\mathbf a_\mu (\mathbf{k})^\dagger]. 
\end{equation}

Let us define:
\be
(\eta,\xi)=\int d\mu(\mathbf{k})\, \tilde \eta_\mu^*(\mathbf k)\, \tilde\xi^\mu(\mathbf k) \label{innerprod}
\ee
and the photon coherent state\footnote{We have slightly generalized the definition of the inner product~\eqref{innerprod} in a natural way to allow inner products with creation and annihilation operators.}~\cite{Sudarshan, Glauber, Kibble1,Kibble2,Kibble3, Kibble4} 
\begin{equation}
|\xi\rangle=e^{(\mathbf a^\dagger,\, \xi)- (\mathbf a,\, \xi)} |0\rangle = e^{\ii A(\xi)}|0\rangle. \label{coherentstate}
\end{equation}
Then
\begin{equation}
\label{sigmafour}
\langle \xi| e^{\ii A(\varphi)} |\xi\rangle=\langle 0| e^{\ii [A(\varphi)+ \Phi_\xi(\varphi)\mathbb{I}]} |0\rangle,
\end{equation}
where 
\begin{equation}
\label{sigmafive}
\Phi_\xi(\varphi)= \sigma( \xi, \varphi).
\end{equation}
Thus (\ref{weyl3}) is a representation built on a generalized coherent state.

On the Fock space we can define the photon number $\mathbf N$  and electromagnetic Hamiltonian  $\mathbf H$ operators as
\be
\mathbf N=\int  \dd\mu(k) \mathbf a^\dagger(\mathbf k) \mathbf a(\mathbf k) \ ; \ \mathbf H=\int  \dd\mu(k)  |\mathbf k|\mathbf (a^\dagger(\mathbf k)\mathbf a(\mathbf k)+\textstyle \frac12).
\ee

If $|\xi\rangle\in \mathcal{H}$  defines an ordinary coherent state, so that it has a finite expectation values for $\mathbf N$ and $\mathbf H$, then the representation is 
equivalent to the one on Fock space. There are however states which belong to the Fock space, and to the domain of $\mathbf H$, but not to the domain of $\mathbf N$. {One can construct sequences  of states that have as extremely low energies $\langle \mathbf H\rangle$ as required, but for which the mean value $\langle\mathbf N \rangle$ of $\mathbf N$ diverges.} Physically these states correspond to  infinities of soft photons. When two  $\langle\mathbf N \rangle$'s differ by infinity, the twisted Weyl representations defined by those  states are  \emph{neither} unitarily equivalent among themselves nor  equivalent to the  original  representation based on the Fock  vacuum $|0\rangle$. They lead to alternative {\it ground states}.

Let us  consider the current 
\begin{equation}
\label{sigmasix}
J_\mu(x)=q \int_{-\infty}^{\infty} d \tau \delta^4(x-x(\tau)) \frac{d x_\mu}{d\tau},
\end{equation}
where for $x(\tau)$ we take the trajectory of a uniform motion of the charge:
 \begin{equation}
\label{ten}
x(\tau)= \frac{p}{m} \tau,
\end{equation}
$p$
being the four-momentum  of the {\it in} state and $q$ the total charge \cite{Roepstorff, balkurqueivaidya,balsolo}.  {At time $t=0$, because of the presence of the charged particle, the system is in the {\sl in} dressed state
  \begin{equation}
\label{eleven}
|J\rangle=\exp \ii\left(q\int_{-\infty}^{0}d t \, \frac{p^\mu}{m} A_\mu\left(\textstyle \frac{p}{m} t \right)\right)|0\rangle.
\end{equation}}

{
The appearance of the extra phase factor is due to the change of the electromagnetic ground state
due to the presence of the charged particle. The Gauss law gets modified
in the presence of charged matter (\ref{dressedttgauss}). Equation (\ref{eleven}), as the solution of lowest energy  goes back at least to Dirac \cite[Eq.\ (31)]{Dirac}. A more recent reference, closer to the spirit of this paper is~\cite[Eq.\ (4.6)]{Roepstorff}.
From the scattering theory viewpoint  the fact that only the retarded fields of  the charged particle 
appear in the dressing factor (\ref{eleven}) is due to the fact that we are considering the ``in'' state in the S matrix formalism.   Moreover, in the semiclassical approximation to the S matrix of full  quantum electrodynamics  where the electron becomes classical the same dressing factor reappears  in the eikonal approximation (see eq. (2.12) of 
   Ref. \cite{Zwanziger}). This argument provides another confirmation that (\ref{eleven}) is the right solution of of Gauss law
   which corresponds to the ground state of the electromagnetic field in the background of a classical charged particle with uniform motion. 
}

The state~\eqref{eleven}
is in fact a linear combination of multi-photon entangled states,
  \bea
|J\rangle&=&|0\rangle + \ii\left(q\int_{-\infty}^{0}d t \, \frac{p^\mu}{m} A_\mu\left(\textstyle \frac{p}{m} t \right)\right)|0\rangle\nonumber\\
&&+\frac{\ii^2}{2} \left(q\int_{-\infty}^{0}d t \, \frac{p^\mu}{m} A_\mu\left(\textstyle \frac{p}{m} t \right)\right)^{\otimes 2}|0\rangle + \dots \label{eleven2}
\eea
The 2-photon state  is an entangled state because it is the sum over several 2-photon states
involved in the integral on $t$. The same happens for higher order terms: the photons generated by 
the charged particle are highly entangled.

 It is known \cite{Roepstorff, balkurqueivaidya,balsolo} that the expectation value of the photon number operator $\mathbf N$ diverges for the coherent state (\ref{eleven}), but the energy remains finite. Thus,  $|J\rangle$ is not in the domain $\mathcal{D}(\mathbf N)$ of the number operator $\mathbf  N$.
We conclude that the infrared-dressed {\it in} state does also not belong to ${\cal D}(\mathbf N)$,
 in spite of the fact that it has a finite energy  $\langle J|\mathbf H |J \rangle$ and thus  $|J\rangle\in \mathcal{D}(H) $.
 
The fact that the state $|J\rangle$ is not in the domain of number operator means that even if repeated measurements of $\mathbf  N$ could give finite results, they do not converge to a mean.
All these ground states  can be heuristically characterized by the presence of an infinity of photons of very low energies. 
In this sense the infrared sector of the theory exhibits  a very rich structure. 

{{The ground state is however a dynamical entity in the presence of the flying particle. One can therefore generalize~\eqref{eleven} to
\be
\label{Jt}
|J\rangle_t=\exp \ii\left(q\int_{-\infty}^{t}d \tau \, \frac{p^\mu}{m} A_\mu\left(\textstyle \frac{p}{m} \tau \right)\right)|0\rangle,
\end{equation}
explicitly showing  that the new ground state is time dependent. 

A gauge invariant observable which is affected by this time variation due to the effect of the moving charge is the energy density of the photon cloud. This is an ultraviolet divergent quantity, like the vacuum energy, because all frequency modes of the cloud of photons contribute to the energy in the same amount. However, in practice the sensitivity of the photodetectors is limited to a range of frequencies 
$k_0<|\mathbf{k}|<k_0+\kappa$. The restriction of the energy density of the cloud of photons to that range of frequencies  is given by
\be
\label{density}
\!\!\!\mathcal{E}_{J}^\kappa(\mathbf x,t)\!=\!_t\!\!\langle J|T_{00}^\kappa(\mathbf{x})|J\rangle_t \!-\!\langle 0|T_{00}(\mathbf{x})|0\rangle\!=\! { \frac{\, q^2\,}{8\pi }}  \kappa (\kappa\!-\!2\,k_0) \textstyle\theta\left(\frac{|\mathbf{p}|}{m} t\!-\!\frac{\mathbf{x}\cdot \mathbf{p}}{{|\mathbf{p}|}}\right) \!
\delta^{(2)}\! \left(\mathbf{x}\!-\!\frac{\mathbf{x}\cdot \mathbf{p}}{{\mathbf{p}}^2}\mathbf{p}\right)\!, \! \end{equation}
where $\theta(s)$ is the Heaviside function which vanishes for negative values of $s$ and is $1$ for positive values of $s$. The argument of the delta function is transverse to $\mathbf{p}$, hence the 2-dimensional delta-function.
The non-trivial variation in time of $\mathcal{E}(\mathbf x,t)$  is the smoking gun of the presence of charged particle in the space.
Notice  that the energy density (\ref{density}) does not vanish along the trajectory of the charge for any  time after the passage of the particle, unlike what happens  in classical electrodynamics.  Thus in the quantum case the {\it memory } of the passage of a charged particle is preserved at any later time. Another characteristic of the photon cloud is the white nature of its transverse energy spectrum, i.e.  the energy contributions of photons of the cloud propagating in transverse directions are independent of their frequencies.

The evident analogy of this effect with the gravitational case might help in understanding the black hole information paradox ~\cite{GomezPanchenko,HawkingStromingerPerry, GomezLetschka, Carneyetal,Calucci,StroZhi,LaddhaSen}.

But we should still check that the theory preserves gauge invariance, namely that the new ground state is still physically acceptable. The rest of the paper is devoted to this verification.
}}

\section{Equations of motion as constraints \label{se:review}}

In \cite{EOMASCON}, we generalised the  Gauss law for pure electromagnetic field to a covariant
Gauss law $G(\eta)$, which depends on  rapidly decreasing smearing space-time one-forms\footnote{The smearing functions are usually taken to be of compact support. In~\cite{EOMASCON} we needed a different behaviour for the Fourier transform of the functions. In this paper we will not discuss these technical issues and assume that the proper limits have been taken. This issue, however, requires further scrutiny.}: 
\begin{equation}
\label{gauss}
G(\eta)=\int d^4x\, \partial^\lambda\, F_{\lambda\mu}(\eta)(x) A^\mu(x),\quad  F_{\lambda\mu}(\eta)=\partial_\lambda \eta_\mu
-\partial_\mu \eta_\lambda, \qquad \eta_\mu\in C^\infty_0(\mathbb{R}^4).
\end{equation}

The quantum vector states $|\psi\rangle_\gamma$ of the free electromagnetic field   compatible with the free field equations of motion  are annihilated by the covariant Gauss law operator $G(\eta)$:
\begin{equation}
\label{gauss1}
G(\eta)|\psi\rangle_\gamma=0.
\end{equation}

The observables are generated by using the gauge invariant smeared fields $A(\varphi)$ as in~\eqref{smearA}.
The constraints   (\ref{gauss1}) are first class, i.e. $ [G(\eta),G(\xi)]=0$,
 as shown by using of the causal propagator
\begin{equation}
[A_\mu(x),A_\nu(y)]=\, \eta_{\mu\nu} D(x-y),
\end{equation}
and integration by parts.

The operators  $G(\eta)$ generate spacetime dependent gauge transformations. Using~\eqref{measure} and partial integrations, we have:
 \begin{equation}
\label{gauss5}
\!\![G(\eta),A_\mu(x)]\!=\!\!  \int\!\! d^4y \, \partial^\lambda F_{\lambda\mu}(\eta)(y) D(y-x)\!=\!\!
-\partial_\mu \int \!\!d^4y \, (\partial^\lambda \eta_\lambda)(y) D(y-x)\!:=\! \partial_\mu \Lambda(x).
\end{equation}
which defines a gauge transformation with gauge function
 \begin{equation}
\label{gauss5a}
\Lambda(x)\!=\! - \int d^4y \,  (\partial^\lambda \eta_\lambda)(y) D(y-x) .
\end{equation}

In the presence of charged matter, the analysis  of gauge invariance slightly changes. In fact the classical equations of motion now read
 \begin{equation}
\label{cgauss}
\partial^\nu F_{\nu \mu}= J_\mu.
\end{equation}
The dressing with   smearing test forms $\eta$ gives
 \begin{equation}
\label{dressedcgauss}
\int d^4x\,\eta^\mu\partial^\nu F_{\nu \mu}(x)= \int d^4x\,\eta^\mu J_\mu(x) = \frac{q} {m}
{p^\mu}\int_{-\infty}^\infty \, d\tau\,\eta_\mu\left(\frac{p}{m} \tau\right) \end{equation}
with $J_\mu$ as in~\eqref{sigmasix}.
{Although gauge transformations  only transform  gauge fields
and not the matter currents, the dressed equations of motion
acquire a non-trivial contribution from the charged sector  as in (\ref{dressedcgauss}).
It implies that  in the presence of charged matter, the generator of gauge transformations
  in the quantum Hilbert space  is given by
 \begin{equation}
\label{dressedtgauss}
G_J(\eta)= G(\eta)-
\int d^4 x\, \eta^\mu J_\mu  \qquad \eta_\mu\in C^\infty_0(\mathbb{R}^4)
 \end{equation}
instead of equation of (\ref{gauss}). The gauge condition on quantum states is
 \begin{equation}
\label{dressedttgauss}
G_J(\eta) |\psi \rangle=0.
 \end{equation}
The equation remains valid if we consider $J_t$ defined in~\eqref{Jt}.}

\section{{Gauge invariance of the ground state of  moving charges} \label{se:passing}}
We start with Minkowski space 
with the photon quantized in the twisted Fock representation induced by the charged particle. 
That  twisted  representation has a typical  {cyclic} vector, i.e. 
the representation is generated from 
the entangled dressed state
  \begin{equation}
|J_t\rangle= e^{\ii A(J_t)} | 0\rangle_\gamma,
\label{dressing}
\end{equation}
with $J_t$ given by~\eqref{Jt} and $ | 0\rangle_\gamma$ being the photon  ground state.
There are many of such representations as each different momentum of the charged particle $p$ gives a different  domain of the Hilbert space. 
{
Each of these twisted ground states generates a different representation of the algebra of local 
observables $\mathbf{A(\varphi)}$. The different representations associated to different momenta define inequivalent superselection sectors of the algebra of gauge invariant local observables. (\ref{smearA}). This phenomenon has been known for a long time~\cite{Eriksson, Kibble1, Kibble2, Kibble3, Kibble4}.
One simple argument is the following: since  $\partial_\mu\varphi^\mu=0$,
\be
[G(\eta),A(\phi)]=0,  \qquad \eta_\mu\in C^\infty_0(\mathbb{R}^4),
\ee
which implies that the different superselection sectors generated by the dressed states are characterized by the eigenvalues of $G(\eta)$: 
\be
G(\eta) A(\phi) | J_0(p)\rangle=   A(\phi) G(\eta) | J_0(p)\rangle=  \frac{q} {m}
{p^\mu}\int_{-\infty}^0 \, d\tau\,\eta_\mu\left(\frac{p}{m} \tau\right) A(\phi)  | J_0(p)\rangle.
\ee
But  for any pair of momenta $p, p'$ of the charged particle, there always exists an $ \eta$ such that 
the two eigenvalues 
\be
{p^\mu}\int_{-\infty}^0 \, d\tau\,\eta_\mu\left(\frac{p}{m} \tau\right), \, {p'^\mu}\int_{-\infty}^0 \, d\tau\,\eta_\mu\left(\frac{p'}{m} \tau\right) 
\ee
are different, which implies that the corresponding twisted sectors  are orthogonal. Moreover,
\be
\langle J_0(p')| A(\phi) | J_0(p)\rangle = 0, \ \ {\rm whenever}\ \  p\neq p',
\ee
which proves the superselected character of the different sectors. 
(See \cite{balnair} for an alternative derivation).
We fix $p$  and the associated twisted Fock representation. 
}

 This state is acted on by observables $\mathbf{A(\varphi)}$ on spacetime. 
We can ask the following question: what happens when we restrict this state to observables which are defined only in a limited causally connected portion of spacetime\footnote{For definiteness one may think of a Rindler wedge $W=\{ x\, \in {\mathbb R}^4; |\mathbf x|> |x_0|\}$.} $W$ and look at the expectation value of $G_J(\eta_{_W})$ with $\eta_{_W}$ being supported in $W$?

The calculation uses only the  dressed Fock space nature of the photon state. Hence consider the ground state expectation value $\langle 0| e^{-I A(J_t)}\, G_{J_t}(\eta_{_W})\,e^{i A(J_t)} |0\rangle$,
suppressing the charged particle state. We find
\bea
\label{fourtops}
\langle 0| e^{-\ii A(J_t)}\, G_{J_t}(\eta_{_W})\,e^{\ii A(J_t)} |0\rangle&=&   _\gamma\langle 0| \, G(\eta_{_W})\,|0\rangle_\gamma -\frac{q}{m} p_\mu \eta_{_W}^\mu\left(\frac pm t\right)
\nonumber\\ && -  \ \ii \int_{-\infty}^{t}d \tau\,  \partial_\tau \int d^4 y \, (\partial\cdot \eta_{_W})(y) D(y-x(\tau)).
\eea
The first term vanishes because of the Gauss law constraint for the photon ground state~(\ref{gauss1}). The second term also vanishes for points outside the support  domain of $\eta_W$. The
third term can be explicitly calculated, the result is $i\Lambda(\textstyle\frac{p}{m}t)$, where $\Lambda$ was defined
in (\ref{gauss5a}). Thus,
\bea
\label{fourtopss}
\langle 0| e^{-i A(J_t)}\, G_{J_t}(\eta_{W})\,e^{i A(J_t)} |0\rangle=  i\Lambda(\textstyle\frac{p}{m}t).
\eea
This relation can be proven noticing that
\bea
\int_{-\infty}^{t}d \tau\,  \partial_\tau \int &&\!\!\!\!\!\!\!\!\!\!\!\!\!\!d^4 y \, (\partial\cdot \eta_{_W})(y) D(y-x(\tau))\\&=& \int d^4 y \, (\partial\cdot \eta_{_W})(y) D(y-\textstyle\frac{p}{m}t)- \int d^4 y \, (\partial\cdot \eta_{_W})(y) D(y-\infty)\nonumber \\
&=&\Lambda(-\infty)-\Lambda(\textstyle\frac{p}{m}t)=-\Lambda(\textstyle\frac{p}{m}t)
\eea
Since the support of $\eta_W$ is in $W$ and $(\vec x,t)\in W'$,  $\Lambda(\textstyle\frac{p}{m}t)$  also vanishes, which proves the gauge invariance of the dressed (\ref{dressing})
state.
The result is not surprising since one would expect the vanishing because of the naked gauge condition 
\be
G(\eta_W)\,e^{-\ii A(J_t)} |J_t\rangle=G(\eta_W)|0\rangle_\gamma=0
\ee
Moreover the equations of motion look all the way to $\partial W'$ as claimed.

 The particle has left the region $W$ so that its support is now completely outside of it. Still the ground state that it has left in its wake, for time going to infinity, is
\be
|J_\infty\rangle=\lim_{t\to\infty}e^{-i A(J_t)}|0\rangle
\ee
This ground state is different from the one we started with. The information of the passing particle is encoded in the new entangled state of photons. In this sense they feel the  {\sl scent} of the particle even after it has left the region. This also happens in classical electrodynamics where the effect of the particle is {\sl retarded}. However, the essential difference in the
quantum case is that the ground state remains entangled for all times and does not depend on a {\sl simple} retarded time.
Moreover, it does not restore the photon Fock  vacuum  at infinite time.

\section{Discussion}
{We argued that the ground state structure  of a limited region of spacetime changes with the passage of a charged particle. Although our discussion was in quantum field theory, and did not introduce gravity (not even as curved background), the photon entanglement will be present also in curved spacetime, possibly with an even richer structure. The mechanism might help in understanding the black hole information paradox. This connects to  ``gravitational memory''~\cite{GomezPanchenko,HawkingStromingerPerry, GomezLetschka, Carneyetal}, as we already commented in the introduction. The role played by soft gravitons~\cite{Calucci,StroZhi,LaddhaSen} is not different from the one played by photons here.  This means that if, for example, upon leaving the region $W$, the particle falls into a black hole, the information is preserved in its scent, in an analogous way. This would involve precising the considerations of this paper to the interacting case.  One would have to introduce a gravitational background at least, and consider other effects, for example decoherence due to bremsstrahlung~\cite{Petruccione1,Petruccione2}. This we leave for another day.}
For the analogous gravitational case there are some proposals for an indirect detection, related with black holes, for example see~\cite{Wang,Lasky} 

Another issue is whether we can gather the information about the particle that has traversed the region $W$. Apart from the fact that it is of course unrealistic that just one particle crosses a region, the drastic change of the way we describe the particle, from a trajectory in phase space to a change in the entanglement of a very large number of photons, render this basically impossible.  It is like asking what happens to the information contained in a book if you burn it in a fireplace. The answer to this apparent paradox is that the smoke pattern of a printed book is different from the one of the same book with blank pages, or a different text. This does not mean that you can read books sitting outside your home watching the smoke! (In any case, burning books is never a good idea). It is conceivable that in some controlled cases, with small regions and strong fields and entanglement, a phenomenon of this kind can be verified. {Likewise an analysis of this kind in the presence of the horizon of a black hole can give a novel perspective for the information paradox.} But this would require further work.

\subsection*{Acknowledgements}
The work of M.A. is partially supported by Spanish MINECO/FEDER grant 
FPA2015-65745-P and DGA-FSE grant 2015-E24/2. 
F.L. and M.A. acknowledge the support of the COST action QSPACE.
 F.L. acknowledges the INFN the Iniziativa Specifica GeoSymQFT 
 and Spanish MINECO under project MDM-2014-0369 of ICCUB (Unidad de Excelencia `Maria de Maeztu').
 G.M. would like to thank the support provided by the Santander/UC3M Excellence Chair Programme 2016/2017.

\end{document}